# Effect of shielding gas composition and welding speed on autogenous welds of unalloyed tungsten plates

G. Marinelli*[,a], F. Martina[a], S. Ganguly[a], S. Williams[a]

[a]Welding Engineering and Laser Processing Centre, College Road, Cranfield University, Cranfield, MK43 0AL, UK

*Corresponding Author. E-mail address: g.marinelli@cranfield.ac.uk (G. Marinelli)

**Abstract**

Tungsten usually exhibits poor weldability and marked brittleness at room temperature. This cause tungsten welds to be affected by the evolution of cracks along the weld bead, which can be eliminated by using a pre-heating step to reduce thermal straining. In this study, based on the tungsten inert gas welding process, a working envelope, focussed on varying welding speed and five different shielding gas mixtures of argon and helium, has been defined with the view of producing crack-free autogenous welds. The bead appearance and the microstructure of the different welds were correlated to the welding parameters, whose main effects have been analysed. Welding defects such as humping occurred when using gas mixtures with relatively low content of helium, and when using relatively high welding speeds. Crack-free autogenous welds have been produced without pre-heating when using a high content of helium and relatively low welding speeds. Thus, this study has demonstrated that a helium-rich shielding gas is required for welding thick tungsten plates. Moreover, the low thermal shock induced by the process, coupled with the purity of the tungsten plates used, strongly contributed to avoid the occurrence of any crack.

**Keywords**

TIG welding, Tungsten, Microstructure, Shielding Gas Composition, Humping

## 1. Introduction

For many years, refractory metals and their alloys have attracted the attention of numerous researchers due to their unique properties and their potential utilisations [1–3]. In fact, this class of materials is principally characterised by high melting point, good strength over a large temperature range, high thermal conductivity and low thermal expansion [4]. For these reasons, the refractories are suitable to manufacture components for high-temperature applications, f.i. parts for future nuclear fusion reactors [5–7].

Among the refractory metals, tungsten has shown the greatest potential to be the most capable and suitable for structural and confinement applications in the nuclear sector [6]. One of the challenges for its industrial utilisation is the development of an effective joining technology capable of ensuring and preserving high structural integrity [8]. The main factors that limit the weldability of this metal are the high melting point and density, the marked brittleness at low temperature, the high reactivity over a wide temperature range, the detrimental effect of impurities on the plasticity and the development of large residual stresses upon cooling [9,10].

A critical review with regards to the suitable joining technologies for refractory metals was conducted by Scott and Knowlson [9] and Olson et al. [10]. With tungsten inert gas (TIG) welding, they reported that the joining must be performed in a high-purity inert environment, and that all the surfaces to be welded should be properly cleaned to rid of contaminants [9,10]. Furthermore, the components to be joined need to be fixed for minimum restraint and pre-heated, preferably above the ductile-brittle transition temperature (DBTT), in order to overcome tungsten's limited plasticity at room temperature [9,10].

The manufacturing route through which tungsten components are manufactured also impacts the weldability. Parts produced via arc-casting (A-C) claimed greater fabricability and higher purity, when compared to components manufactured via powder metallurgy (PM) or chemical vapour deposition (CVD) routes [9,11,12]. In fact, when joining tungsten components manufactured via PM and CVD, porosity usually develops [10].

Lessmann and Gold [12] have previously investigated the weldability of unalloyed A-C tungsten and unalloyed PM tungsten sheets using TIG and electron beam (EB) welding. Welding speeds between 1.5 mm/s and 20 mm/s have been used in combination with no pre-heating, 250° C pre-heating and 700° C pre-heating. The plates had a thickness of around 0.9 mm. They reported that the pronouncedly higher thermal shock from the EB welding (compared to TIG welding) caused dramatic failures, including cracks transverse to the weld seam and delamination of adjacent base metal [12]. The occurrence of cracks was reported for both manufacturing routes, with the additional issue of porosity for the PM tungsten sheets. A lower process thermal shock seemed to have a benefit on the welded structure, with a lower occurrence of cracking. In particular, this phenomenon is evident when TIG welding plates pre-heated to 700° C [12].

Cole et al. [13] also evaluated the TIG weldability of tungsten plates produced by A-C, PM and CVD. The welds were performed on 1.5-mm-thick plates in an inert atmosphere using a welding current of 350 A, argon as shielding gas and a welding speed of around 4.0 mm/s. Welds on A-C material were free from porosity, unlike the welds performed on PM plates which were usually porous, particularly along the fusion line [13]. The occurrence of cracks was eliminated for single pass welds when using pre-heating at around 150° C.

Additional studies of welding of CVD, PM and vacuum A-C tungsten sheets were reported by Farrell et al. [14]. They performed bead-on-plates welds on 1.5-mm-thick plates using TIG welding in order to examine the occurrence of porosity and cracks. The process employed the utilisation of complete inert atmosphere and pre-heating at about 150° C. Pure argon was used as shielding gas, with welding current between 310 A and 350 A, and a welding speed of around 4.0 mm/s. Cracks were observed within the weld bead of PM and vacuum A-C tungsten, while CVD tungsten plates presented cracks within their heat affected zone. Their occurrence has been identified as intergranular hot cracks caused by the growth and coalescence of grain-boundary pores [14].

Currently, there are few published pieces of research that describe the process development and the weldment microstructure for unalloyed tungsten using TIG welding. These investigations also mainly refer to the welding of thin tungsten plates. Furthermore, no studies have been found to discuss the effect of different gas mixtures for tungsten welding. This study principally aims at understanding the correct process parameters for tungsten welding when using TIG welding process, and to define an operating window for this process. Furthermore, one of the objectives of this study is to produce defect-free welds without using pre-heating. The relationship between weldment morphology and two fundamental welding parameters (welding speed and shielding gas composition) was established, providing an analysis of the main causes of resulting microstructure and welding defects.

## 2. Experimental Procedure

The autogenous welds were performed on commercially available unalloyed tungsten plates. Each of the plates, manufactured by PM, was 100 mm in length, 90 mm in width and 5 mm in thickness. Their chemical composition is listed in **Table 1**. Plates were ground and rinsed with acetone to eliminate surface contamination before welding. Furthermore, high-purity argon and high-purity helium were used as shielding gases.

**Table 1**
Elemental composition (wt. %) of tungsten plates.

|  | W | Mo | Ta | Ti | V | Cr | Fe | C | N | O | K |
|---|---|---|---|---|---|---|---|---|---|---|---|
| W Plates | 99.99 | <0.05 | <0.05 | <0.05 | <0.05 | <0.05 | <0.05 | <10 ppm | <10 ppm | <50 ppm | <10 ppm |

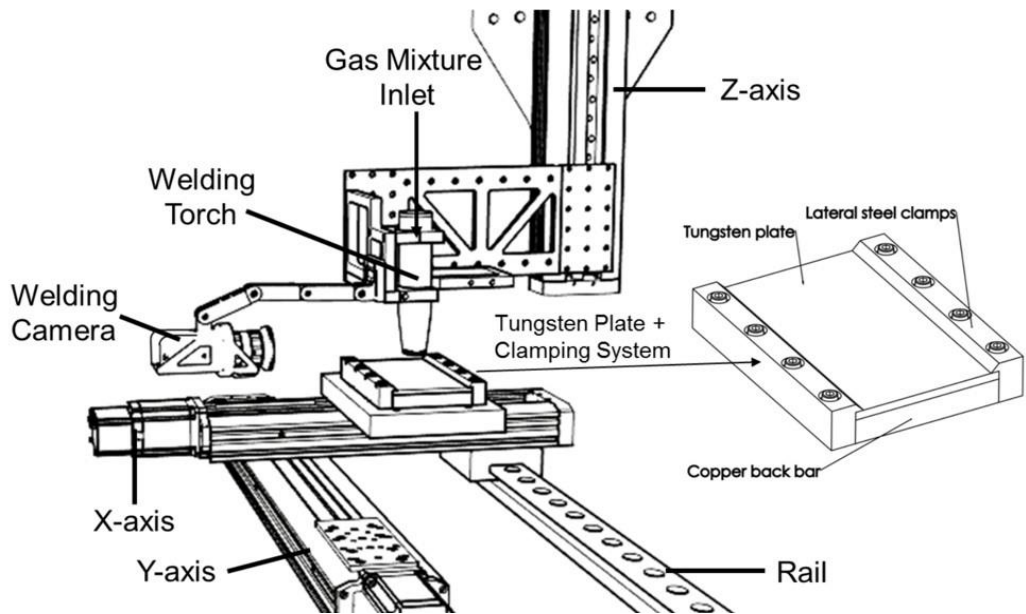

**Fig. 1**: Schematisation of the experimental set-up used to produce the tungsten weldments.

**Fig. 1** shows the layout of the welding apparatus used in this study. A conventional TIG torch and a power supply were used. The heat source and the clamping system were attached to three linear motorized high-load stages assembled in the XYZ configuration. The tungsten plates were held in position using the clamping system also shown in **Fig. 1**.

Every screw of the clamping system was set using the same torque of 25 N to have an identical resulting clamping force. A gas mixing system that employed two separate gas channels was used to mix argon and helium in a controlled fashion, so as to study the effect of different shielding gas compositions. The outlet of the gas mixing system was directly connected to the welding torch. All the welds were performed in an inert environment with a concentration of oxygen of ~100 ppm. The process was monitored using a welding camera; the electrical signals of arc voltage and current were recorded via an AMV 5000.

The experimental matrix is conveniently reported in **Table 2.** Twenty 80-mm-welds were performed on unalloyed tungsten plates. Four welds were performed next to each other, on the same plate. Whilst shielding gas composition (SGC) and welding travel speed (TS) were varied, other welding parameters were kept constant, throughout the experiment, to reduce the number of variables. These are listed as follows:

- Welding current (I) = 350 A
- Electrode-to-workpiece distance = 3.5 mm;
- Tungsten electrode tip angle = 45°;
- Tungsten electrode diameter = 3.6 mm;
- Torch angle= 90°.

**Table 2**
Experimental matrix showing the changes in the levels of welding travel speed (TS) and shielding gas composition (SGC).

| Sample name | TS [mm/s] | SGC He [%] | SGC Ar [%] |
|---|---|---|---|
| 1 | 1 | 0 | 100 |
| 2 | 2 | 0 | 100 |
| 3 | 4 | 0 | 100 |
| 4 | 6 | 0 | 100 |
| 5 | 1 | 25 | 75 |
| 6 | 2 | 25 | 75 |
| 7 | 4 | 25 | 75 |
| 8 | 6 | 25 | 75 |
| 9 | 1 | 50 | 50 |
| 10 | 2 | 50 | 50 |
| 11 | 4 | 50 | 50 |
| 12 | 6 | 50 | 50 |
| 13 | 1 | 75 | 25 |
| 14 | 2 | 75 | 25 |
| 15 | 4 | 75 | 25 |
| 16 | 6 | 75 | 25 |
| 17 | 1 | 100 | 0 |
| 18 | 2 | 100 | 0 |
| 19 | 4 | 100 | 0 |
| 20 | 6 | 100 | 0 |

Three sections were extracted from each weldment as illustrated in **Fig. 2** and analysed. All the metallographic samples were ground and polished using paper discs of silicon carbide and then etched using the Murakami's reagent to study the occurrence of defects, the microstructure and geometrical features of the melted regions.

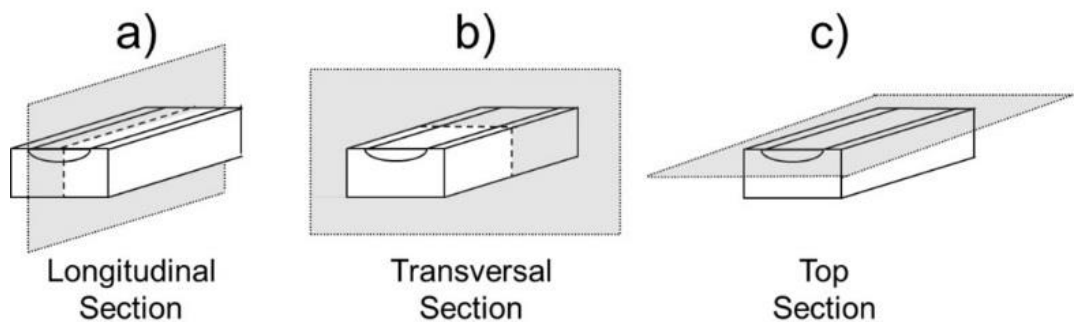

**Fig. 2**: Schematic of the planes used to analyse the microstructure of the samples produced indicated as (a) longitudinal cross-section, (b) transversal cross-section and (c) top section.

## 3. Results and Discussion

*3.1. Operating window and bead morphology*

**Fig. 3a** shows the operating window explored in this study. The black dots shown in the graph represent the single experiments, according to the set of parameters reported in **Table 2**. The boundaries of each region were drawn based on interpolated lines between the testing points.

Four main weld bead morphologies were identified, for different combinations of TS and SGC. These have been nominated for simplicity: "No Weld" (**Fig. 3e**), "Humping" (**Fig. 3d**), "Transient" (**Fig. 3c**) and "Good Weld" (**Fig. 3b**). The occurrence of each case has been illustrated as a coloured area within the operative window of **Fig. 3a**.

In general, the No Weld morphology occurred for relatively low-power cases (low helium content) and it was characterised by the absence of clear and visible melted surface. At these specific levels of power, the development of a liquid pool did not occur during the process. Regarding the Humping morphology, clear melted surfaces (reminiscent of an established melt pool) were visible, but they showed periodical series of crests followed by depressions. The Transient morphology presented a visible transition from the humping regime to a smooth weld, within the same bead. In this case, the humps gradually decrease their periodicity, eventually disappearing into a regular weld profile towards the end of a weld. Finally, the Good Weld profiles were limited to high content of helium (from 75% to 100% in composition), and low welding speeds, particularly 1 mm/s and 2 mm/s. Noteworthy was the absence of welding defects and solidification cracks for these welds.

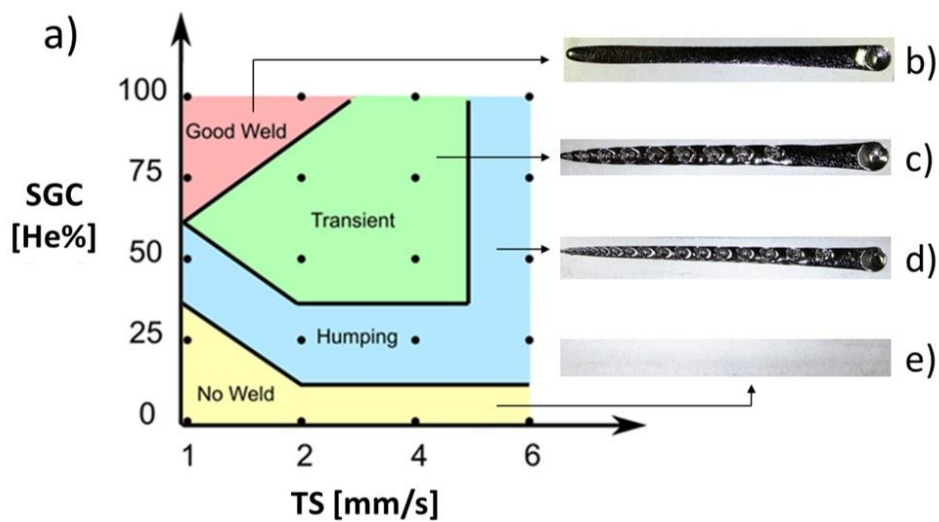

**Fig. 3**: (a) operating window with experimented combinations (the black dots) of shielding gas composition (balance: argon) and travel speed; (b) weld bead morphology of a Good Weld; (c) Transient; (d) Humping; (e) No Weld.

Each morphology was explained by the process heat input and the arc characteristics given by each specific mixture used. In particular, it is well known that the heat input per unit length for the TIG welding process ($H_{in}$) is proportional to the welding current (I), the arc voltage drop (V) and the welding travel speed (TS), according to the following **Eq.(1)** [15]:

$$H_{in} = \eta \ \frac{I*V}{TS} \qquad (1)$$

Where η is the process efficiency, which was not measured in this study. Whilst the TS was directly varied as one of the main parameters of the investigation, the voltage changed depending on the content of helium within the gas mixture. In fact, a helium arc has a higher voltage when compared to an argon arc, due to helium's higher ionization potential [16]. In **Table 3**, the average voltage values, measured during the welding process, are reported for each gas mixture; it is possible to see that the voltage increased as the content of helium increased, in agreement with the above.

In TIG, the shielding gas composition can also affect the total heat flux going from the arc column to the workpiece [17,18]. The influence of the physical properties of the shielding gas on the electric arc characteristics has been already investigated extensively [19]. In particular, helium is characterised by a lower electrical conductivity, a higher specific heat and a higher thermal conductivity with respect to argon [19].

**Table 3**
Average measured voltage for the different shielding gas compositions (I=350A).

| SGC [He%] | Voltage [V] |
|---|---|
| 0 | 14.5 |
| 25 | 16 |
| 50 | 18.5 |
| 75 | 20 |
| 100 | 21.5 |

The mixtures of argon and helium have intermediate properties to those of the pure gases [17,19]. The addition of a helium fraction to the argon arc increases the thermal conductivity and the arc constriction, thus promoting an increased heat flux and current density to the workpiece [18]. Due to the high melting point and the relatively high thermal conductivity of tungsten, a 100% helium arc and relatively low welding speeds were essential to the formation and the development of a regular weld pool. Conversely, higher welding speeds and lower helium content led to the formation of humping defects, mainly due to the reduction in heat input, heat flux and current density. These are covered in the following section.

*3.2. Humping formation*

**Fig. 4** shows the microstructure of one of the samples produced and characterised by the occurrence of humping, defined as the occurrence of an alternation of crests (humps) and valleys (undercut) on the top of the weld [23]. In particular, **Fig. 4a** refers to the longitudinal cross-section and **Fig. 4b** refers to the top section of the weld bead (please refer to **Fig. 2** for the reference system). When analysing **Fig. 4a**, it is possible to notice that the average melting depth remained constant along the weld, despite the occurrence of humping. The average melting depth and the average humping height resulted to be both equal to 0.7 mm, with respect to the top of the plate.

Moreover, the weld width and the humping pitch were found to be constant (**Fig. 4b**) and equal to 3.8 mm and 6.0 mm, respectively. The humping mainly occurred for a low content of helium over the entire range of welding speeds explored, and for high welding speed when a higher content of helium was employed. This means that the heat input was sufficiently high to melt the tungsten plate, but the liquid metal was pushed downwards and forced to flow towards the rear of the weld pool, where it solidified prematurely.

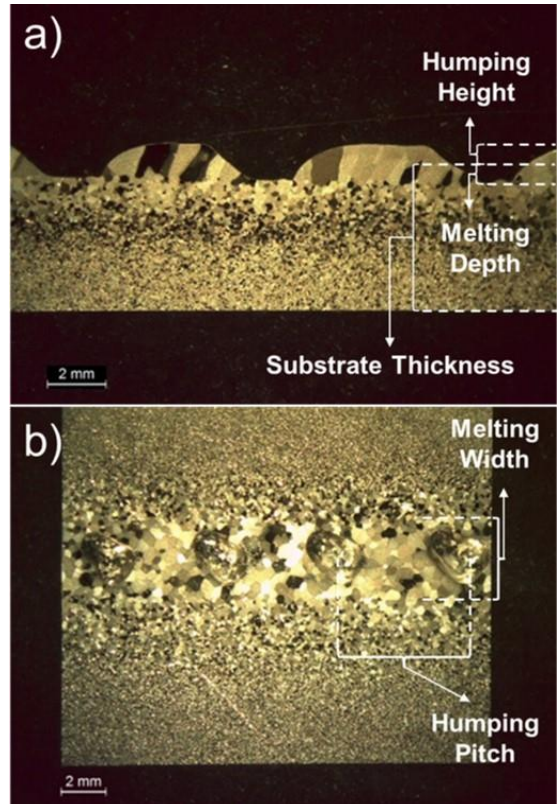

**Fig. 4**: Longitudinal cross-section (a) and top section (b) of the weld performed using a shielding gas composition of 50% of helium and 1 mm/s of welding speed.

The constant pitch shows that humping happened occurred regularly along the weld. This defect has already been reported for high-TS and high-current welds [15,20–22]. In particular, increasing TS beyond a certain critical limit leads to the formation of humping. This critical speed values decreases as the welding current increases [24]. Furthermore, as the composition of the shielding gas influences the total arc pressure, this could contribute to the formation of humping [24]. Generally, at high welding currents, the arc pressure causes a marked depression on the weld pool surface [25], forming a thin layer of liquid metal, called the gouging region [20]. The humps start to form when the arc pressure enlarges this region beyond the heat influence of the arc column causing a premature solidification of the thin liquid layer [21].

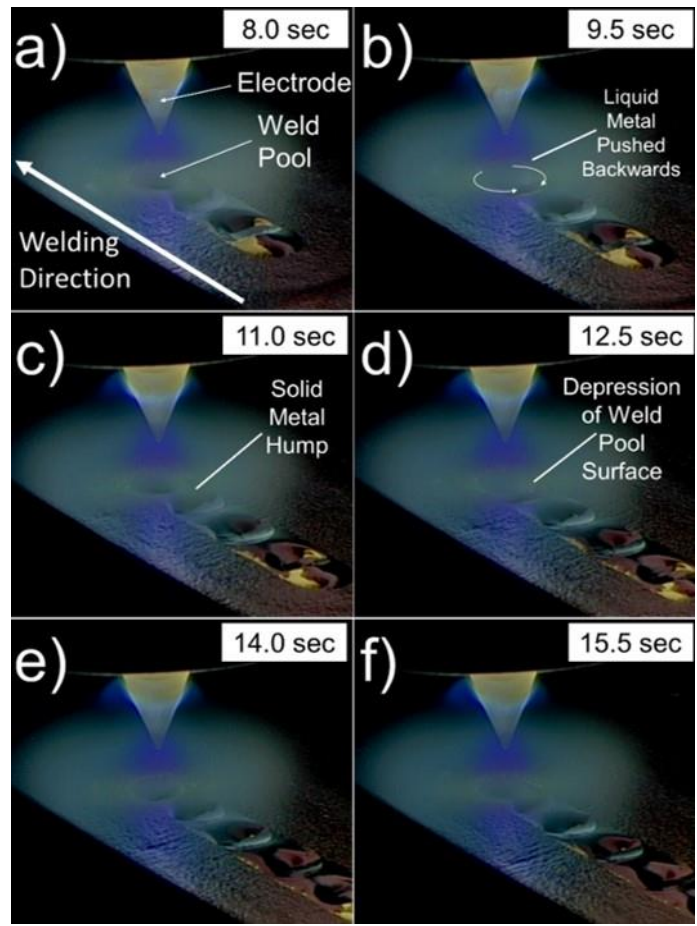

**Fig. 5**: Time-resolved images of the weld performed using a shielding gas composition of 50% of helium and 1 mm/s of welding speed.

**Fig. 5** shows a series of time-resolved images of those welds performed using a welding speed of 1 mm/s and a shielding arc composition of 50% of argon and 50% of helium. In **Fig. 5a,** it is possible to notice the presence of a large depression on the weld pool surface right underneath the arc column. Here, the weld pool has been reduced to a thin liquid layer that flows rapidly from the front to the rear of the liquid pool (**Fig. 5b**) through lateral channels [20]. For these reasons, the liquid experienced an upwards deflection, forming the hump (**Fig. 5c**) and also, the lateral channels solidified as they were extended to a region away from the arc column [23]. In this situation, the depression located under the arc column was not filled back with liquid metal and so the undercutting occurred (**Fig. 5d**) [15,20]. Thus, insufficient welding heat input causes humping as the displaced metal by the arc pressure solidifies before completion of the flow. This effect is enhanced at relatively high arc pressures and high welding speeds.

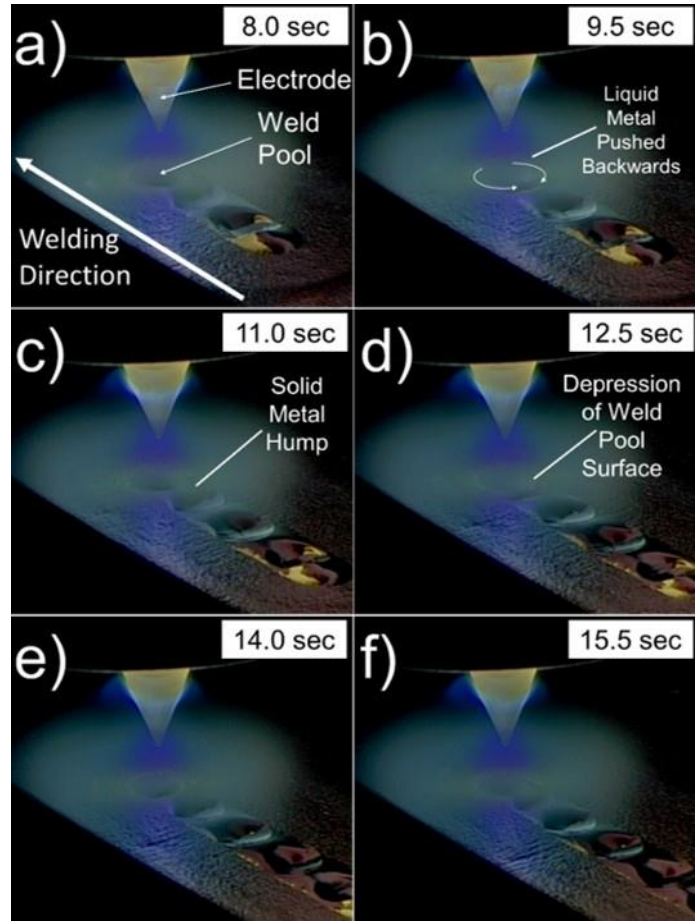

**Fig. 5**: Time-resolved images of the weld performed using a shielding gas composition of 50% of helium and 1 mm/s of welding speed.

*3.3. Transient welding state*

**Fig. 6** shows the evolution of the weld pool when using 75% of helium within the gas mixture and a welding speed of 2 mm/s. It is possible to see how, starting with the formation of humping (**Fig. 6a**), the pitch of the humps increased, after 12.0 seconds, as the weld progressed (**Fig. 6b**). This occurred until a transition to a regular weld profile was obtained achieved toward the end of the weld (**Fig. 6c**), where the weld bead did not show defects any longer. This scenario represents a transient welding state which was mainly caused by three factors: the considerable variation in thermal gradient caused by the decreasing in thermal conductivity of tungsten within the same weld, the relatively high content of helium and the relatively low travel speed.

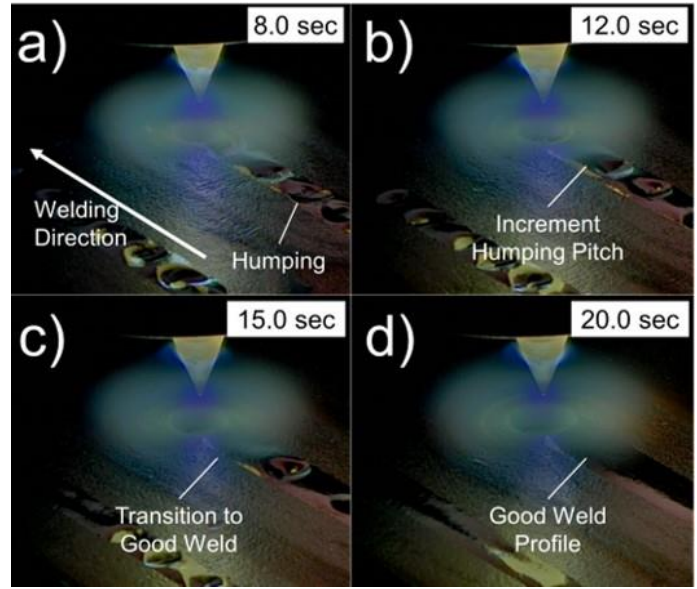

**Fig. 6**: Time-resolved images of the weld performed using a shielding gas composition of 75% of helium and 2 mm/s of welding speed.

In fact, it is already known that the thermal conductivity of tungsten decreases when the temperature increases [26–28]. In addition, the shear stress exerted by the helium arc onto the weld pool was previously measured to be smaller than argon and so that resulted to be for the arc pressure. [17,18,29]. Furthermore, it has been already seen that the frequency of the humping decreases for relatively low welding speed, as a result of an increased heat input per unit length [30]. For these reasons, as the weld progressed, the thermal gradient was reduced, due to a constantly decreasing thermal conductivity. Consequently, the process' melting efficiency increased towards the end of the weld. The progressively larger liquid volume was not completely displaced away from the arc column, as during humping, due to the higher metallostatic pressure and the smaller arc pressure of the 75% helium arc [20,24].

*3.4. Good weld morphology*

**Fig. 7** shows the weld pool evolution over time when 100% helium shielding gas was used, with a welding speed of 1 mm/s. Clearly, the weld pool edges were well-defined since the beginning of the process (**Fig. 7a**). Indeed, no humping was registered for the entire weld, as shown in **Fig. 7b** and **Fig. 7c.**

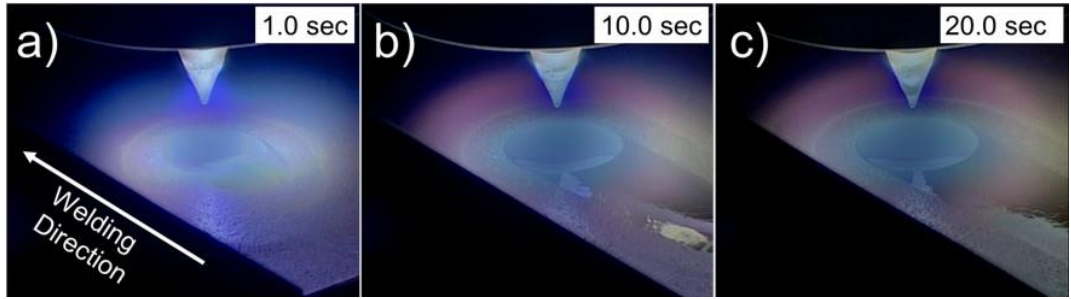

**Fig. 7**: Time-resolved images of the weld performed using a shielding gas composition of 100% of helium and 1 mm/s of welding speed.

The microstructure of the Good Weld achieved using 100% helium arc and 1 mm/s of welding speed is reported in **Fig. 8**. In particular, **Fig. 8a** shows the transversal cross-section, **Fig. 8b** shows the longitudinal cross-section and **Fig. 8c** shows the top section (as defined in **Fig. 2**). No clear fusion line has been found at the interface between the melted region and heat affected zone. The fusion boundary has been therefore chosen based on the sudden change in the grain dimensions. This has been used to characterise the average weld width and depth, measured at 6.3 mm and 1.45 mm, respectively.

The melted material seemed to be characterised by numerous elongated grains extending from the bottom toward the top of the weld pool. Each grain possibly nucleated from the smaller grains found in the parent material, and grew epitaxially toward the top of the weld pool.

In this study, a relatively large amount of power of 7 kW has been used to achieve the Good Weld condition. For this reason, the absence of cracks for the autogenous weld performed in this study can be explained by the extremely low level of impurities (**Table 1**), the limited restraint from the clamping system (**Fig. 1**) and the relatively low process-induced thermal shock. In fact, it has already been reported that elements like carbon, oxygen and nitrogen tend to increase the brittleness of the welded material mainly due to their limited solubility within tungsten [9,10]. In particular, they tend to segregate at the grain boundaries promoting the formation of brittle compounds such as carbides and oxides [31,32]. Furthermore, the relatively low plasticity and the high modulus of elasticity of tungsten lead to the development of a considerable level of residual stresses, possibly causing the evolution of cracks upon cooling [10]. This is also influenced by the main process parameters as welding speed and heat input. In fact, slower welding speeds and higher heat inputs lead to a reduction of the process-induced thermal shock [10]. Thus, the cracks in tungsten welding are mainly caused by the high level of impurities and high process thermal shock, which causes high levels of strain.

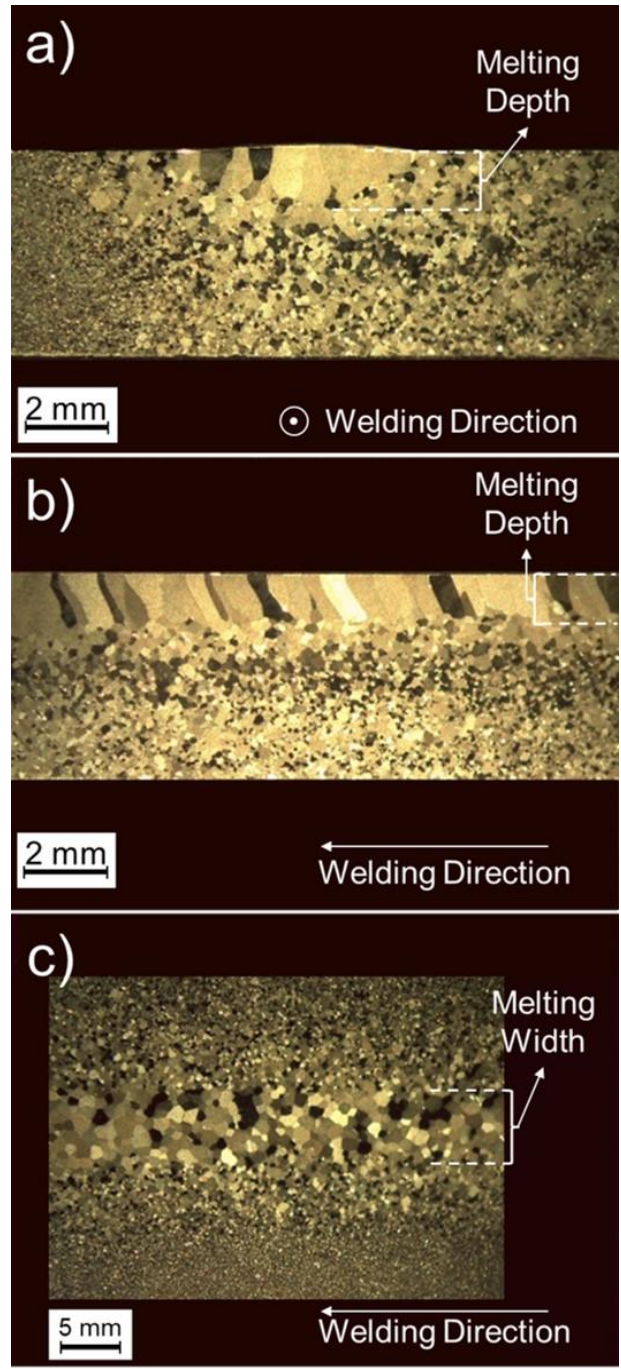

**Fig. 8**: Transversal cross-section (a), longitudinal cross-section (b) and top section (c) of the weld performed using a shielding gas composition of 100% of helium and 1 mm/s of welding speed.

## 4. Conclusion

In this research, the influences of the welding speed and the shielding gas composition on the autogenous welds of tungsten plates have been investigated. Furthermore, the occurrence of humping and the transient welding regime have been also discussed with relation to the arc characteristic and the thermal properties of tungsten. The summary of the main finding is reported below:

- The addition of helium to the shielding gas was necessary to achieve effective melting of the 5-mm-thick plates using a welding current of 350 A. A high content of helium is necessary to develop a smooth and regular weld pool without any defects, for the range of welding speeds studied;

- The occurrence of welding defects as humping was reported for high welding speed or low content of helium within the shielding gas. The humping defects were caused by force extended by the arc pressure on the liquid metal, which caused the premature solidification of the weld pool. The occurrence of this type of defect was associated with low heat input;

- The transient welding state from the humping to the good weld profile has been caused by the overall reduction of the thermal gradient, which is associated with the reduction in thermal conductivity of tungsten with temperature. The increased liquid metal volume and the high content of helium also contributed to reduce the occurrence of the humping;

- The absence of cracks within the good weld was mainly due to low thermal shock, low level of restraining from the clamping system and the low level of impurities of the tungsten plates.

This research has shown that it is possible to obtain defect-free welds of tungsten plates without the use of pre-heating. The design of a proper heat source is fundamental to preserve weldments integrity. A low process thermal shock and high thermal energy were a fundamental requirement to avoid the occurrence and evolution of cracks within tungsten welds. However, increased productivity could be achieved by using pre-heating, which will lead to a further reduction of the thermal gradient and thermal shock.

## Acknowledgement

The authors wish to acknowledge financial support from the AMAZE Project, which was co-funded by the European Commission in the 7th Framework Programme (contract FP7-2012-NMP-ICT-FoF-313781), by the European Space Agency and by the individual partner organisations.